\documentclass[12pt,twoside]{article}
\usepackage{epsf}
\usepackage{amsfonts} 
\usepackage{amsmath} 
\usepackage[latin2]{inputenc} 
\usepackage[T1]{fontenc} 
\usepackage{epsfig} 
 
\newlength{\dinwidth}                          
\newlength{\dinmargin} 
\def\beq{\begin{equation}}  
\def\eeq{\end{equation}}  
\def\bea{\begin{eqnarray}}   
\def\eea{\end{eqnarray}}   
\def\bq{\begin{quote}}   
\def\eq{\end{quote}}   
\def\bi{\begin{itemize}}   
\def\ei{\end{itemize}}   
\def\beqa{\begin{eqnarray}}   
\def\eeqa{\end{eqnarray}}   
\def\be{\begin{enumerate}}   
\def\ee{\end{enumerate}}   
\def\beq{\begin{equation}}   
\def\eeq{\end{equation}}   
\def\bi{\begin{itemize}} 
\def\ei{\end{itemize}}

\def\r2{\sqrt{2}}   
 
\def\bi{\begin{itemize}}   
\def\ei{\end{itemize}}

\def\beq{\begin{equation}}  
\def\eeq{\end{equation}}  
\def\bea{\begin{eqnarray}}   
\def\eea{\end{eqnarray}}   
\def\bq{\begin{quote}}   
\def\eq{\end{quote}}   
\def\bi{\begin{itemize}}   
\def\ei{\end{itemize}}   
\def\beqa{\begin{eqnarray}}   
\def\eeqa{\end{eqnarray}}   
\def\be{\begin{enumerate}}   
\def\ee{\end{enumerate}}   
\def\beq{\begin{equation}}   
\def\eeq{\end{equation}}   
\def\bi{\begin{itemize}} 
\def\ei{\end{itemize}}  
\def\bc{\begin{center}} 
\def\ec{\end{center}}

\def\r2{\sqrt{2}}   
 
\def\bi{\begin{itemize}}   
\def\ei{\end{itemize}}

\def\beq{\begin{equation}}  
\def\eeq{\end{equation}}  
\def\bea{\begin{eqnarray}}   
\def\eea{\end{eqnarray}}   
\def\bq{\begin{quote}}   
\def\eq{\end{quote}}   
\def\bi{\begin{itemize}}   
\def\ei{\end{itemize}}   
\def\beqa{\begin{eqnarray}}   
\def\eeqa{\end{eqnarray}}   
\def\be{\begin{enumerate}}   
\def\ee{\end{enumerate}}   
\def\beq{\begin{equation}}   
\def\eeq{\end{equation}}   
\def\bi{\begin{itemize}} 
\def\ei{\end{itemize}}

\def\r2{\sqrt{2}}   
  
\def\bi{\begin{itemize}}   
\def\ei{\end{itemize}}

\def\bc{\begin{center}} 
\def\ec{\end{center}}   
 
\DeclareMathAlphabet{\scr}{U}{rsfs}{m}{n}  
\begin{document} 
\pagestyle{empty}  
\begin{flushright}   CERN-TH/2001-260 
\end{flushright}  
\vskip 2cm  
\begin{center}  
{\Huge  
Boundary Terms in Brane Worlds} 
  
\vspace*{5mm} \vspace*{1cm}   
\end{center}  
\vspace*{5mm} \noindent  
\vskip 0.5cm  
\centerline{\bf Zygmunt Lalak${}^{1,2}$ 
and Rados\l aw Matyszkiewicz${}^2$}  
\vskip 1cm  
\centerline{\em ${}^{1}$Theory Division, CERN}  
\centerline{\em CH-1211 Geneva 23, Switzerland}  
\vskip 0.3cm  
\centerline{\em ${}^{2}$Institute of Theoretical Physics}  
\centerline{\em University of Warsaw, Poland}  
\vskip 2cm  
  
\centerline{\bf Abstract}  
\noindent We clarify the relation between orbifold and interval pictures in 5d brane worlds. We establish this correspondence for $Z_2$-even and  
$Z_2$-odd orbifold fields. In the interval picture Gibbons--Hawking terms  
are necessary to fulfill consistency conditions. We show how the  
brane world consistency conditions arise  
in the interval picture. We apply the procedure to the situation where  
the transverse dimension is terminated by naked singularities. In particular,  
we find the boundary terms needed when the naive vacuum action is infinite.  
\vskip3cm  
\begin{flushleft}October 2001\end{flushleft}
\newpage  
\pagestyle{plain}   
\section{Introduction}  
In studies of higher dimensional field theories  
with branes, it is convenient to  model compact extra dimensions  
with the help of orbifolds \cite{hw,randal1,randal2}.  
Branes are usually put at orbifold fixed  
points, and various matter and gauge sectors are   
localized on their world volumes. Yet another gauge sector may propagate, together with gravity and moduli fields, in the bulk of an orbifold, interacting  
with branes. In many cases, e.g. when one considers $AdS_5$ throats that form 
 near stacks of $D3$ branes in string theory, it is appropriate to  
restrict a discussion  
to the simplest class of five-dimensional field theories.
It has been noticed long ago \cite{hw} that there are two ways  
of looking at field theory in such a geometry. One possibility is 
 to notice that a circle divided by $Z_2$ is actually a line segment  
with the ends at the fixed points of the action of $Z_2$. In this picture  
the fixed points are the ends of the world, beyond which the fields living 
in the bulk cannot propagate. The second possibility, sometimes called
the `upstairs picture', consists in working on the whole circle $S^1$, but  
defining in a non-trivial way the action of the $Z_2$ on the fields.  
The second possibility turned out to be the most convenient, especially  
in the case of supersymmetric theories. There, the fields fall into one of two classes: the $Z_2$-even fields, which have massless modes in their Kaluza--Klein  
expansion and are continuous across the branes, and the $Z_2$-odd fields, which  
have non-zero modes only and may have finite discontinuities when crossing the  
branes.  
Of course it would be wrong to say, as one may naively think, that only  
the even modes survive when going over to the interval picture.  
The point is that, away from the branes, the bulk does not know whether it is  
investigated in the first or in the second picture; it must be therefore  
exactly the same in both. Hence the difference between various fields must  
lie in the way they couple to the branes. The analogy with electrodynamics  
may be helpful. The electromagnetic scalar potential is continuous across 
a surface with a non-zero charge density; however, it suffers a discontinuity  
on a surface with a non-zero density of an electric dipole moment.  
The difference between the two situations lies precisely in  the form of the coupling of the potential to the surface. In what follows we shall  
discuss in some detail the relation between the upstairs and interval pictures 
for 
the $Z_2$-even, $Z_2$-odd and gravitational fields. It comes down to the determination of the boundary terms (couplings) for various fields in the interval  
picture, given the field theory defined in a natural way on the circle.  
Alternatively, one may start with the action on the interval augmented by  
Neumann or Dirichlet boundary conditions on the boundaries, depending on 
whether we work with the field that is supposed to have a zero mode or not; 
given boundary couplings, which enforce required boundary conditions, one can promote them to the upstairs picture.  
Among other cases we  discuss a situation where the boundary is located at  
the position of a naked singularity (including   
one at coordinate infinity).  
We find out that, boundary terms are necessary in the interval picture,  
to satisfy consistency conditions that are  
analogous to those known from the upstairs picture \cite{Ellwanger:2000pq,
Forste:2000ps,Forste:2000ft,sumrules}.  
An interesting bonus for establishing the complete equivalence between the  
two pictures is the possibility of adding together two orbifold theories,  
for example of forming a supergravity that describes two supergravities  
living in two separate throats near two separate stacks of D3 branes and  
connected through the common Planck brane \cite{Dimopoulos:2001ui}. Adding two intervals with a common end is rather obvious, whereas adding models living on  
two  circles is not immediately clear. 
\section{$Z_2 $-even and $Z_2 $-odd scalar fields in the interval picture}
We begin with the case of a $Z_2$-even scalar field, that is the one which  
is allowed to have a zero mode, in the interval picture.   
As usual, we take a five-dimensional manifold that consists of a four-dimensional spacetime multiplied by an interval: $ x_{M}=(x_{\mu},y)\in \Omega=M^{4}\times \langle 0,L \rangle$. 
 Two four-dimensional branes, which couple to the induced 4d  
metric tensor, are located at $y=0$ and $y=L$ and endowed with an induced metric 
        \begin{equation} 
         g_{\mu\nu}^{0}(x_{\rho})\equiv G_{\mu\nu}(x_{\rho},y=0)\ ,\;\;\; 
         g_{\mu\nu}^{L}(x_{\rho})\equiv G_{\mu\nu}(x_{\rho},y=L)\ , 
        \end{equation}  
        where $G_{MN}$ denotes a five-dimensional metric tensor, and $M,N=
0,1,...,5$. We take the usual choice $G_{\mu 5}=0$. 
The classical action describing the scalar field in 5d reads:  
        \begin{equation} \label{dzialanie} 
        S = -\frac{1}{2}\int d^{4}x\int^{L}_{0} dy \sqrt{-G}\left(G^{MN}\partial_{M}\Phi\partial_{N}\Phi+V(\Phi)\right)\ , 
        \end{equation} 
        where $V(\Phi)$ is a scalar potential. 
Its variation  with respect to $\Phi$ after integrating by parts is: 
        \begin{eqnarray} 
        &\delta S = &-\int d^{4}x\int^{L}_{0} dy \sqrt{-G}\left(-G^{MN}\partial_{M}\partial_{N}\Phi-\frac{1}{\sqrt{-G}}\partial_{M}\left(\sqrt{-G}G^{MN}\right)\partial_{N}\Phi+\frac{1}{2}\frac{\partial V(\Phi)}{\partial \Phi}\right)\delta\Phi \nonumber \\ && -\int d^{4}x\sqrt{-G}G^{55}\partial_{5}\Phi\delta\Phi\Big|_{y=L}+\int d^{4}x\sqrt{-G}G^{55}\partial_{5}\Phi\delta\Phi\Big|_{y=0}  . 
        \end{eqnarray} 
Since the interval is finite, and branes are, presumably, physical objects,  
it is natural to allow for non-vanishing variations $\delta \Phi$ at the endpoints. This leads to Neuman boundary conditions for the field  $\Phi$. 
To obtain classical equations of motion   
       we must modify the action (\ref{dzialanie}) by adding other boundary terms: 
        \begin{equation} \label{czlonyskalar} 
        S'=S+\int d^{4}x\sqrt{-g^{L}}V_{L}(\Phi)\Big|_{y=L}+\int d^{4}x\sqrt{-g^{0}}V_{0}(\Phi)\Big|_{y=0}, 
        \end{equation} 
        with additional terms satisfying the conditions: 
	\begin{equation} \label{warunkiskalarne} 
	\sqrt{-G}G^{55}\partial_{5}\Phi\Big|_{y=0}=-\sqrt{-g^{0}}\frac{\partial V_{0}}{\partial\Phi}\Big|_{y=0}\ ,\qquad\sqrt{-G}G^{55}\partial_{5}\Phi\Big|_{y=L}=\sqrt{-g^{L}}\frac{\partial V_{L}}{\partial\Phi}\Big|_{y=L}\ . 
	\end{equation}  
One can think of an orbifold $S^1/Z_2$ as a sum of two intervals with ends  
at the fixed points. Physics on the two spaces is correlated by the action of  
the orbifold symmetry $Z_2$. We shall take the action (\ref{czlonyskalar})  
for each interval, impose $Z_2$ symmetry, compare the resulting equation of motion with those directly derived in the orbifold picture, and read off the relation between the boundary terms in both pictures.  
Let us denote the boundary terms on the intervals as  
        \begin{equation}  
        S_{+}= \int d^{4}x\sqrt{-g^{0}}\frac{1}{2}V_{0}^{+}\Phi\Bigr|_{y=0}+\int d^{4}x\sqrt{-g^{\pi r_{c}}}\frac{1}{2}V_{\pi}^{+}\Phi\Bigr|_{y=\pi r_{c}}\ , 
        \end{equation}  
        for $y\in (0,\pi r_{c})$, and 
        \begin{equation}  
        S_{-}= \int d^{4}x\sqrt{-g^{0}}\frac{1}{2}V_{0}^{-}\Phi\Bigr|_{y=0}+\int d^{4}x\sqrt{-g^{-\pi r_{c}}}\frac{1}{2}V_{\pi}^{-}\Phi\Bigr|_{y=-\pi r_{c}} 
        \end{equation}  
        for $y\in (-\pi r_{c},0)$, and   
        impose $Z_2$ parity $\Phi(y)=\Phi(-y)$.   
        This implies 
        \begin{equation}  
        V_{0}^{+}=V_{0}^{-}\ ,\qquad V_{\pi}^{+}=V_{\pi}^{-}\ . 
        \end{equation}   
The corresponding action on the orbifold is 
        \begin{eqnarray}  
        &S=&-\frac{1}{2}\int d^{4}x\oint_{-\pi r_{c}}^{\pi r_{c}}dy\sqrt{-G}\left(G^{MN}\partial_{M}\Phi\partial_{N}\Phi+V(\Phi)\right)+\nonumber\\&&+\int d^{4}x\sqrt{-g^{0}}V_{0}\Phi\Bigr|_{y=0}+\int d^{4}x\sqrt{-g^{\pi r_{c}}}V_{\pi}\Phi\Bigr|_{y=\pi r_{c}}\ , \end{eqnarray} 
        where $y\in\langle-\pi r_{c},\pi r_{c}\rangle$, 
 and the equations of motion are  
        \begin{eqnarray} \label{skalarorbifold} 
&&       \sqrt{-G} G^{MN} \partial_{M} \partial_{N} \Phi+\partial_{M}  
\left( \sqrt{-G} G^{MN} \right ) \partial_{N} \Phi= \nonumber \\ &=&\sqrt{-G} \frac{1}{2} \frac{\partial V(\Phi)}{\partial \Phi}- 
\sqrt{-g^{0}}V_{0} \delta(y)-\sqrt{-g^{\pi r_{c}}} V_{\pi}  
\delta(y-\pi r_{c}) .  
        \end{eqnarray}  
 
We note that there are no additional boundary terms from the  
derivation of the equation of motion, since the integral of a full divergence of any field vanishes on an orbifold, even if the field is not continuous.  
A $Z_2$-even scalar field is a function of an absolute value of $y$.  
Hence one finds 
	\begin{eqnarray}  
        &\Phi=\Phi(|y|)\ ,\qquad\Phi'=\epsilon(y)\Phi'(|y|)\ ,&\nonumber\\&\Phi''=2\left(\delta(y)-\delta(y-\pi r_{c})\right)\Phi'(|y|)+\Phi''(|y|)\ ,& 
        \end{eqnarray}  
        and equation (\ref{skalarorbifold}) leads to: 
        \begin{equation}  
        \sqrt{-G}\partial_{y}\Phi\Bigr|_{y=0^{+}}=-\sqrt{-g^{0}}\frac{V_{0}}{2}\ ,\qquad\sqrt{-G}\partial_{y}\Phi\Bigr|_{y=\pi r_{c}^{-}}=\sqrt{-g^{\pi r_{c}}}\frac{V_{\pi}}{2}\ . 
        \end{equation}   
It is clear that we have an identification between two intervals and an orbifold  if  
        \begin{equation}  
        V_{0}=V_{0}^{+}=V_{0}^{-}\ ,\qquad V_{\pi}=V_{\pi}^{+}=V_{\pi}^{-}\ . 
        \end{equation}   
 
In the next step let us perform the same analysis for a field, which has no zero mode on the orbifold, that is for the field which is $Z_2$-odd.  
The derivative of an odd field with respect to $y$ is  
$Z_2$-even. Hence, one can   
 construct an action where such a field couples to branes through its  
first derivative 
\begin{equation} \label{nieparzystedzialanie} 
        S= \int_{0}^{L}dy\int d^{4}x\sqrt{-G}\left(-\frac{1}{2}G^{MN}\partial_{M}\Phi\partial_{N}\Phi+\frac{1}{\sqrt{-G}}\partial_{M}\left(\sqrt{-G}G^{MN}\Phi\partial_{N}\Phi\right)-\frac{1}{2}V(\Phi)\right)\ . 
\end{equation} 
This is the analogue of the coupling of the electric potential to the electric dipoles on the brane, which appeared before in the 5d constructions of  
\cite{Mirabelli:1998aj,Falkowski:2000er,Falkowski:2001yq}.  
This action differs from the action  (\ref{dzialanie}) by a full divergence  
of a vector field. On manifolds without boundary, such as an orbifold,  
these additional terms vanish. On an interval we obtain non-zero  
boundary contributions: 
        \begin{equation}  
        S= -\frac{1}{2}\int_{0}^{L}dy\int d^{4}x\sqrt{-G}\left(G^{MN}\partial_{M}\Phi\partial_{N}\Phi+V(\Phi)\right)+\int d^{4}x\sqrt{-G}G^{55}\Phi\partial_{5}\Phi\Bigr|_{0}^{L}\ . 
        \end{equation}  
The variation of this action  with respect to the scalar field gives 
        \begin{eqnarray} \label{wariacjaniepskal} 
        &\delta S&= \int_{0}^{L}dy\int d^{4}x\sqrt{-G}\left(G^{MN}\partial_{N}\partial_{M}\Phi+\frac{1}{\sqrt{-G}}\partial_{M}\left(\sqrt{-G}G^{MN}\right)\partial_{N}\Phi-\frac{1}{2}\frac{\partial V(\Phi)}{\partial \Phi}\right)\delta\Phi+ \nonumber\\&&+\int d^{4}x\sqrt{-G}G^{55}\Phi\delta(\partial_{5}\Phi)\Bigr|_{0}^{L}\ . 
        \end{eqnarray} 
        If we do not insist that $\delta(\partial_{5}\Phi)$ vanishes on the boundaries, we need to add other terms in (\ref{nieparzystedzialanie}), the variation of which cancels the boundary terms in (\ref{wariacjaniepskal}): 
        \begin{equation} \label{wlasciwyoddskalar} 
        S'=S+\int d^{4}x\sqrt{-G}G^{55}V_{0}\partial_{5}\Phi\Bigr|_{y=0}+\int d^{4}x\sqrt{-G}G^{55}V_{L}\partial_{5}\Phi\Bigr|_{y=L}\ , 
        \end{equation}  
        where $\Phi(0)=V_{0}\ ,\;\Phi(L)=-V_{L}\ $.

To make the relation between orbifold and interval pictures explicit,  
let us consider a $Z_2$-odd scalar field  on an orbifold  
	\begin{equation} \label{dzialaniedlaniep} 
	S=-\frac{1}{2}\oint_{-\pi r_{c}}^{\pi r_{c}}dy\int d^{4}x\sqrt{-G}\left(G^{MN}\partial_{M}\Phi\partial_{N}\Phi+m^{2}\Phi^{2}+V_{0}(y)\Phi+V_{\pi}(y)\Phi\right)\ , 
	\end{equation}  
        where $y\in\langle-\pi r_{c},\pi r_{c}\rangle$. Tensions $V_{0}$, $V_{\pi}$ are proportional to one-point distributions $\delta$ and $\delta'$ at points $y=0$ and $y=\pi r_{c}$ respectively. The variation with respect to the scalar field leads to
        \begin{equation} \label{skalarorbifoldodd}  
        G^{MN}\partial_{M}\partial_{N}\Phi+\frac{1}{\sqrt{-G}}\partial_{M}\left(\sqrt{-G}G^{MN}\right)\partial_{N}\Phi-m^{2}\Phi-\frac{1}{2}V_{0}(y)-\frac{1}{2}V_{\pi}(y)=0 
        \end{equation}   
        in  five-dimensional spacetime. Again, no additional boundary terms  
appear, since the integral over the whole orbifold of a total divergence  
always vanishes.  
        A $Z_2$-odd scalar field may be represented as a function of an  
absolute value of $y$ multiplied  by the antisymmetric step function  
$\epsilon(y)$ (we shall define $\epsilon(0)=0$). With this taken into account we obtain 
        \begin{eqnarray}  
        &\Phi=\epsilon(y)\bar{\Phi}(|y|)\ ,\qquad\Phi'=\bar{\Phi}'(|y|)+2\left(\delta(y)-\delta(y-\pi r_{c})\right)\bar{\Phi}(|y|)\ ,&\nonumber\\& \Phi''=\epsilon(y)\bar{\Phi}''(|y|)+2\left(\delta(y)-\delta(y-\pi r_{c})\right)\epsilon(y)\bar{\Phi}'(|y|)+2\left(\delta'(y)-\delta'(y-\pi r_{c})\right)\bar{\Phi}(|y|)\ ,& 
        \end{eqnarray}  
        and the relation (\ref{skalarorbifoldodd}) leads to 
	\begin{equation}  
	V_{0}(y)=4G^{55}\epsilon(y)\bar{\Phi}'(|y|)\delta(y)+\frac{4}{\sqrt{-G}}\left(\sqrt{-G}G^{55}\right)'\bar{\Phi}(|y|)\delta(y)\ +4G^{55}\bar{\Phi}(|y|)\delta'(y) 
	\end{equation}  
        and 
	\begin{eqnarray} & V_{\pi}(y)=&-4G^{55}\epsilon(y)\bar{\Phi}'(|y|)\delta(y-\pi r_{c})- \frac{4}{\sqrt{-G}}\left(\sqrt{-G}G^{55}\right)'\bar{\Phi}(|y|)\delta(y-\pi r_{c})+\nonumber \\ & & -4G^{55}\bar{\Phi}(|y|)\delta'(y-\pi r_{c}). \end{eqnarray} 
Let us rewrite brane tensions in action (\ref{dzialaniedlaniep}) in such a way that they contain only terms proportional to $\delta$ and no factors of  $\delta'$.  
        A partial integration of the boundary terms in (\ref{dzialaniedlaniep}) leads to: 
        \begin{eqnarray} &  S&\supset\oint_{-\pi r_{c}}^{\pi r_{c}}dy\int d^{4}x\sqrt{-G}\left(-2G^{55}\bar{\Phi}(|y|)\Phi'(y)\delta(y-\pi r_{c})+2G^{55}\bar{\Phi}(|y|)\Phi'(y)\delta(y)\right)= \nonumber \\ & & =-\int d^{4}x\sqrt{-G}G^{55}\Lambda_{\pi}\Phi'(y)\Bigr|_{y=\pi r_{c}}+\int d^{4}x\sqrt{-G}G^{55}\Lambda_{0}\Phi'(y)\Bigr|_{y=0}\ ,\end{eqnarray} 
        where $\Lambda_{0}=2\bar{\Phi}(0)$ and $\Lambda_{\pi}=2\bar{\Phi}(\pi r_{c})$. 
         
It should be noted that the action (\ref{dzialaniedlaniep}) is singular, because of non-vanishing terms proportional to $\delta^{2}$: 
        \begin{equation} 
        S\sim  \oint_{-\pi r_{c}}^{\pi r_{c}}dy\int d^{4}x\sqrt{-G}\left(2G^{55}\bar{\Phi}^{2}(|y|)\delta^{2}(y)-2G^{55}\bar{\Phi}^{2}(|y|)\delta^{2}(y-\pi r_{c})\right)\ . 
        \end{equation}  
       We can eliminate this singularity by adding to the initial action (\ref{dzialaniedlaniep}) new terms, that do not change the equations of motion, 
        \begin{equation}  
        S'=S+\frac{1}{2}\oint_{-\pi r_{c}}^{\pi r_{c}}dy\int d^{4}x\sqrt{-G}\left(-G^{55}\Lambda^{2}_{0}\delta^{2}(y)+G^{55}\Lambda^{2}_{\pi}\delta^{2}(y-\pi r_{c})\right)\ . 
        \end{equation}  
This modified action can be put into a `full square' form  
        \begin{eqnarray} &  S'=&-\frac{1}{2}\oint_{-\pi r_{c}}^{\pi r_{c}}dy\int d^{4}x\sqrt{-G}\left(G^{\mu\nu}\partial_{\mu}\Phi\partial_{\nu}\Phi+m^{2}\Phi^{2}\right)- \nonumber \\ & & -\frac{1}{2}\oint_{-\pi r_{c}}^{\pi r_{c}}dy\int d^{4}x\sqrt{-G}G^{55}\left(\partial_{5}\Phi-\Lambda_{0}\delta(y)+\Lambda_{\pi}\delta(y-\pi r_{c})\right)^{2}\, . \end{eqnarray} 
This is the structure that plays a crucial role in the brane-bulk supersymmetric  
models \cite{hw,Mirabelli:1998aj,Falkowski:2000er,Falkowski:2001yq}.  
\vskip0.3cm 
Let us turn to the interval picture.  
We write a $Z_2$-odd scalar field action (\ref{wlasciwyoddskalar}), on  
two separate intervals   
	\begin{eqnarray} &   S_{+}= &-\frac{1}{2}\int_{0}^{\pi r_{c}}dy\int d^{4}x\sqrt{-G}\left(G^{MN}\partial_{M}\bar{\Phi}\partial_{N}\bar{\Phi}+m^{2}\bar{\Phi}^{2}\right)+\int d^{4}x\sqrt{-G}G^{55}\bar{\Phi}\partial_{5}\bar{\Phi}\Bigr|_{0^{+}}^{\pi r_{c}^{-}}+ \nonumber \\ & & +\int d^{4}x\sqrt{-G}G^{55}V_{0}^{+}\partial_{5}\bar{\Phi}\Bigr|_{y=0^{+}}+\int d^{4}x\sqrt{-G}G^{55}V_{\pi}^{+}\partial_{5}\bar{\Phi}\Bigr|_{y=\pi r_{c}^{-}}\ ,\end{eqnarray} 
        for $y\in (0,\pi r_{c})$, and 
	\begin{eqnarray} &   S_{-}= &-\frac{1}{2}\int^{0}_{-\pi r_{c}}dy\int d^{4}x\sqrt{-G}\left(G^{MN}\partial_{M}\bar{\Phi}\partial_{N}\bar{\Phi}+m^{2}\bar{\Phi}^{2}\right)+\int d^{4}x\sqrt{-G}G^{55}\bar{\Phi}\partial_{5}\bar{\Phi}\Bigr|^{0^{-}}_{-\pi r_{c}^{+}}+\nonumber \\ & & +\int d^{4}x\sqrt{-G}G^{55}V_{0}^{-}\partial_{5}\bar{\Phi}\Bigr|_{y=0^{-}}+\int d^{4}x\sqrt{-G}G^{55}V_{\pi}^{-}\partial_{5}\bar{\Phi}\Bigr|_{y=\pi r_{c}^{+}}\ ,\end{eqnarray} 
        for $y\in (-\pi r_{c},0)$. It is easy to find, following the steps taken in the case of an even field, that we obtain the identification between two intervals and an orbifold  
when 
        \begin{equation}  
        \Lambda_{0}=2V_{0}^{+}=2V_{0}^{-}\ ,\qquad \Lambda_{\pi}=-2V_{\pi}^{+}=-2V_{\pi}^{-}\ , 
        \end{equation}  
        and $\Phi(y)=\epsilon\bar{\Phi}(|y|)$. 
 
We have established the correspondence between boundary terms and couplings  
for fields of definite $Z_2$ parity on $S^1 / Z_2$ and corresponding fields living on the sum of two intervals. Since parity correlates the physics
on the two intervals in a unique way, it is sufficient to work with just one of them. {}From the point of view of the interval, the difference between odd and even fields lies in the way they couple to the branes.  
\vspace{0.3cm} 
\section{Gravity in the interval picture}
To complete the discussion of the relation of brane worlds on   
orbifolds and intervals we need to include gravity. It is well known that,
on the interval, we need to add Gibbons--Hawking boundary terms to recover  
Einstein equations; nevertheless, it is instructive to find the form of the  
required terms from a reasoning analogous to that used before for scalar fields. The procedure we are going to use was considered earlier  
by Dick \cite{dick} (see also \cite{Silva:2001ys}).  We begin with the classical gravitation action: 
         \begin{equation} \label{dzialanieG} 
        S = \int d^{4}x\int^{L}_{0} dy\sqrt{-G}\left(M^{3}R+{\scr L}\right)+\int d^{4}x \sqrt{-g^{0}}L^{0}\Bigr|_{y=0} +\int d^{4}x \sqrt{-g^{L}}L^{L}\Bigr|_{y=L}\ , 
        \end{equation} 
        where ${\scr L}$ is a Lagrangian of matter fields in the bulk,  
and $L^{0}$, $L^{L}$ describe matter fields on four-dimensional branes located at $y=0$ and $y=L$ respectively. It is well known  
     that the variation with respect to the metric tensor is 
        \begin{eqnarray}\label{fst} &  \delta S& = \int d^{4}x\int^{L}_{0} dy\sqrt{-G}\left[\left(M^{3}R_{MN}-\frac{1}{2}M^{3}RG_{MN}-T_{MN}\right)\delta G^{MN}+M^{3}G^{MN}\delta R_{MN}\right]+ \nonumber \\ & & -\int d^{4}x \sqrt{-g^{0}}T^{0}_{\mu\nu}\delta g_{0}^{\mu\nu}\Bigr|_{y=0} -\int d^{4}x \sqrt{-g^{L}}T^{L}_{\mu\nu}\delta g_{L}^{\mu\nu}\Bigr|_{y=L}\ ,\end{eqnarray} 
        where 
	\begin{equation} \label{definicjaT} 
	 T_{MN}=\frac{1}{2}{\scr L}G_{MN}-\frac{\partial {\scr L}}{\partial G^{MN}}\ ,\quad  T^{0}_{\mu\nu}=\frac{1}{2} L^{0}g^{0}_{\mu\nu}-\frac{\partial L^{0}}{\partial g_{0}^{\mu\nu}}\ ,\quad T^{L}_{\mu\nu}=\frac{1}{2} L^{L}g^{L}_{\mu\nu}-\frac{\partial L^{L}}{\partial g_{L}^{\mu\nu}}\ . 
	\end{equation} 	 
The last term in the first line of (\ref{fst}) can be written as  
        \begin{equation}  
        \int d^{4}x\int^{L}_{0} dyM^{3}\partial_{M}\left(\sqrt{-G}W^{M}\right)=\int d^{4}x\sqrt{-G}M^{3}W^{5}\Big|_{y=L}-\int d^{4}x\sqrt{-G}M^{3}W^{5}\Big|_{y=0}\ , 
        \end{equation}   
where 
        \begin{displaymath}  
        W^{M}=G^{OP}\delta\Gamma^{M}_{OP}-G^{OM}\delta\Gamma^{P}_{OP}\ . 
        \end{displaymath} 
The difference between the case of a scalar field and the present one  
is that the boundary terms contain variations of the metric tensor together with variations of its derivatives. To improve this situation one can start with the modified action          
        \begin{equation} \label{dzialaniepop} 
        S = \int d^{4}x\int^{L}_{0} dy\sqrt{-G}\left(M^{3}F+{\scr L}\right)+\int d^{4}x \sqrt{-g^{0}}L^{0}\Bigr|_{y=0} +\int d^{4}x \sqrt{-g^{L}}L^{L}\Bigr|_{y=L}\ , 
        \end{equation}  
where  
        \begin{displaymath}  
        F=G^{MN}\left(\Gamma^{P}_{MS}\Gamma^{S}_{NP}-\Gamma^{P}_{MN}\Gamma^{S}_{PS}\right)\ . 
        \end{displaymath} 
The modified action gives Einstein equations in the bulk, but  
there appear boundary terms 
	\begin{equation} 
  \int d^{4}x\sqrt{-G}M^{3}W^{5}_{MN}\delta G^{MN}\Big|_{y=L}-\int d^{4}x\sqrt{-G}M^{3}W^{5}_{MN}\delta G^{MN}\Big|_{y=0}\ ,   
\end{equation} 
where 
	\begin{eqnarray} \label{definicjaW} & W^{5}_{MN}&=\frac{1}{2}\left(G_{OM}\partial_{N}G^{O5}+G_{ON}\partial_{M}G^{O5}-G_{OM}G_{PN}\partial^{5}G^{OP}\right)+\nonumber \\ & & +\frac{1}{2}\left(-G_{MN}\partial_{P}G^{P5}-G_{MN}G^{OP}\partial^{5}G_{OP}+G^{OP}\partial_{N}G_{OP}\delta_{M}^{5}\right)\ . \end{eqnarray} 
One needs to take care of the variations of boundary terms. The requirement that these variations vanish leads to the conditions  
        \begin{eqnarray}  \label{warunkinak} & \sqrt{-G}M^{3}W^{5}_{\mu\nu}\Big|_{y=0}=-\sqrt{-g^{0}}T^{0}_{\mu\nu}\Bigr|_{y=0}\ ,& \nonumber \\ & \sqrt{-G}M^{3}W^{5}_{\mu\nu}\Big|_{y=L}=\sqrt{-g^{L}}T^{L}_{\mu\nu}\Bigr|_{y=L}\ , & \end{eqnarray} 
        and 
        \begin{displaymath}  
        W_{55}^{5}\Big|_{y=0}=W_{55}^{5}\Big|_{y=L}=0\ . 
        \end{displaymath} 
Let us check whether these conditions are satisfied in the Randall--Sundrum model, with $ 
        ds^{2} = e^{-2k|y|}\eta_{\mu\nu}dx^{\mu}dx^{\nu} + dy^{2}\ .$ 
Using (\ref{definicjaW}), (\ref{definicjaT}), and counting boundary 
terms twice (because of the orbifold symmetry)  
we obtain 
        \begin{equation}  
        W^{5}_{\mu\nu}=6kG_{\mu\nu}\ , 
        \end{equation} 
        and 
        \begin{equation}  
        T^{hid}_{\mu\nu}=-6kM^{3} g^{hid}_{\mu\nu}\ ,\ \ \ \ T^{vis}_{\mu\nu}=6kM^{3} g^{vis}_{\mu\nu}\ , 
        \end{equation}  
in full agreement with (\ref{warunkinak}). 
       
To make the connection with the standard gravitational action, we note that 
        \begin{equation}  
        R=F+\frac{1}{\sqrt{-G}}\partial_{M}\left(\sqrt{-G}\bar{W}^{M}\right)\ , 
        \end{equation}  
        where  
        \begin{displaymath}  
        \bar{W}^{M}=G^{NS}\Gamma_{NS}^{M}-G^{MN}\Gamma_{NS}^{S}\ . 
        \end{displaymath} 
This leads to the following form of (\ref{dzialaniepop})  
        \begin{eqnarray}\label{czlonki} & S&=\int d^{4}x\int^{L}_{0} dy\sqrt{-G}\left(M^{3}R+{\scr L}\right)-\int d^{4}x\sqrt{-G}M^{3}\bar{W}^{5}\Big|_{y=L}+\int d^{4}x\sqrt{-G}M^{3}\bar{W}^{5}\Big|_{y=0}+\nonumber \\ & &+\int d^{4}x \sqrt{-g^{0}}L^{0}\Bigr|_{y=0} +\int d^{4}x \sqrt{-g^{L}}L^{L}\Bigr|_{y=L}\ . \end{eqnarray} 
As a result, we obtain the standard action augmented by boundary terms, which should be further extended by brane potentials.         
One easily finds out that new terms are equal to the well known Gibbons--Hawking terms: 
        \begin{equation}  
        S_{HG}=-\int_{\partial\Omega}d^{4}x\sqrt{-g}2M^{3}K\ , 
        \end{equation}  
        where 
        \begin{displaymath}  
        K=-g^{\mu\nu}\nabla_{\mu}n_{\nu}=g^{\mu\nu}\Gamma_{\mu\nu}^{S}n_{S}\ ,  
        \end{displaymath} 
        and $n_{M}$ is the unit normal to the surface 
$\partial\Omega$. In the 5d case one has  
        \begin{displaymath}  
        K=-\frac{1}{2}g^{\mu\nu}\partial_{P}G_{\mu\nu}n^{P}\ , 
        \end{displaymath}  
       implying 
        \begin{equation} \label{howki} 
        S_{HG}=\int d^{4}x\sqrt{-g^{L}}M^{3}g_{L}^{\mu\nu}\partial_{5}G_{\mu\nu}\Big|_{y=L}-\int d^{4}x\sqrt{-g^{0}}M^{3}g_{0}^{\mu\nu}\partial_{5}G_{\mu\nu}\Big|_{y=0}\ . 
        \end{equation}  
This is equal to (\ref{czlonki}) after taking into account that 
        \begin{equation} 
	\bar{W}^{5}=G^{NS}\Gamma_{NS}^{5}-G^{5N}\Gamma_{NS}^{S}=-G^{\mu\nu}\partial^{5}G_{\mu\nu}\ .
	\end{equation} 
         
	The RS model on an orbifold should be 
equivalent to the same model on two intervals. In the orbifold picture  
the effective cosmological constant vanishes \cite{randal1}.  
Let us do the calculation in the interval picture. First,  let us  
calculate without the Gibbons--Hawking terms.  
        After inserting the vacuum solution into the action, we obtain  
        \begin{equation}  
        -\Lambda_{eff}=2\int_{0}^{\pi r_{c}}dy\sqrt{-G}e^{-4ky}\left(-20M^{3}k^{2}-\Lambda\right)+V_{0}+V_{\pi}e^{-4k\pi r_{c}}\ , 
        \end{equation}   
        where 
        \begin{equation}  
         V_{0}=-V_{\pi}=12M^{3}k\ ,\qquad\Lambda=-12M^{3}k^{2}\ . 
        \end{equation}  
      The result is non-zero: 
        \begin{equation}  
        -\Lambda_{eff}=16M^{3}k\left(e^{-4k\pi r_{c}}-1\right)\ . 
        \end{equation}  
        Let us now take into account the Gibbons--Hawking terms 
        \begin{equation}  
        \bar{W}^{5}=-G^{\mu\nu}\partial^{5}G_{\mu\nu}=8k\ .  
        \end{equation}       
        This time we obtain the correct result 
        \begin{equation}  
        -\Lambda_{HG}=-16M^{3}k\left(e^{-4k\pi r_{c}}-1\right)\ , 
        \end{equation}  
        \begin{equation}  
        \Lambda_{eff}+\Lambda_{HG}=0\ . 
        \end{equation}  
\vskip0.3cm 
To be more explicit,  
 let us take an action (\ref{czlonki}) defined on two separate intervals:  
	\begin{eqnarray} \label{dzialintgraw}&  S_{+}&= \int d^{4}x\int_{0}^{\pi r_{c}}dy\sqrt{-G}\left(M^{3}R+{\scr L}\right)+\int d^{4}x\sqrt{-G}\left[M^{3}\bar{W}^{5}+\frac{1}{2\sqrt{G_{55}}}L_{0}^{+}\right]_{y=0}+ \nonumber \\ & & +\int d^{4}x\sqrt{-G}\left[-M^{3}\bar{W}^{5}+\frac{1}{2\sqrt{G_{55}}}L_{\pi}^{+}\right]_{y=\pi r_{c}}\ , \end{eqnarray} 
        for $y\in (0,\pi r_{c})$, and 
        \begin{eqnarray} &  S_{-}& = \int d^{4}x\int_{-\pi r_{c}}^{0}dy\sqrt{-G}\left(M^{3}R+{\scr L}\right)+\int d^{4}x\sqrt{-G}\left[-M^{3}\bar{W}^{5}+\frac{1}{2\sqrt{G_{55}}}L_{0}^{-}\right]_{y=0}+ \nonumber \\ & &  +\int d^{4}x\sqrt{-G}\left[M^{3}\bar{W}^{5}+\frac{1}{2\sqrt{G_{55}}}L_{\pi}^{-}\right]_{y=-\pi r_{c}}\ , \end{eqnarray} 
        for $y\in (-\pi r_{c},0)$. 
The variation of the action with respect to the metric tensor gives  
        \begin{equation}  
        \sqrt{-G}\left(R_{MN}-\frac{1}{2}RG_{MN}-\frac{T_{MN}}{M^{3}}\right)=0 , 
        \end{equation}  
        in the bulk, and 
	\begin{eqnarray}\label{warbrz1} &  \sqrt{-G}M^{3}W^{5}_{\mu\nu}\Bigr|_{y=0^{+}}=-\sqrt{-g^{0}}\frac{T_{\mu\nu}^{0^{+}}}{2}\Bigr|_{y=0^{+}}\ ,\qquad \sqrt{-G}M^{3}W^{5}_{\mu\nu}\Bigr|_{y=\pi r_{c}^{-}}=\sqrt{-g^{\pi}}\frac{T_{\mu\nu}^{\pi^{+}}}{2}\Bigr|_{y=\pi r_{c}^{-}}\ , &\nonumber \\ &  \sqrt{-G}M^{3}W^{5}_{\mu\nu}\Bigr|_{y=0^{-}}=\sqrt{-g^{0}}\frac{T_{\mu\nu}^{0^{-}}}{2}\Bigr|_{y=0^{-}}\ ,\qquad \sqrt{-G}M^{3}W^{5}_{\mu\nu}\Bigr|_{y=-\pi r_{c}^{+}}=-\sqrt{-g^{\pi}}\frac{T_{\mu\nu}^{\pi^{-}}}{2}\Bigr|_{y=-\pi r_{c}^{+}}\ , &\end{eqnarray} 
        on the boundary. Taking the natural parity assignments $G_{\mu\nu}(y)=G_{\mu\nu}(-y)$ and $G_{55}(y)=G_{55}(-y)$, we obtain 
        \begin{equation}  
        T_{\mu\nu}^{0^{+}}=T_{\mu\nu}^{0^{-}}\ ,\qquad T_{\mu\nu}^{\pi^{+}}=T_{\mu\nu}^{\pi^{-}}\ . 
        \end{equation}  
On an orbifold the Gibbons--Hawking terms are absent; one is given  
the condition        \begin{equation}  
        \sqrt{-G}M^{3}W^{5}_{\mu\nu}\Bigr|_{y=0^{+}}=-\sqrt{-g^{0}}\frac{T^{0}_{\mu\nu}}{2}\ ,\qquad \sqrt{-G}M^{3}W^{5}_{\mu\nu}\Bigr|_{y=\pi r_{c}^{-}}=\sqrt{-g^{\pi}}\frac{T^{\pi}_{\mu\nu}}{2}\ . 
        \end{equation}   
It follows that the identification of the interval and orbifold pictures  
requires 
        \begin{equation} \label{warunkirownowa}  
        T_{\mu\nu}^{0}=T_{\mu\nu}^{0^{+}}=T_{\mu\nu}^{0^{-}}\ ,\qquad T_{\mu\nu}^{\pi}=T_{\mu\nu}^{\pi^{+}}=T_{\mu\nu}^{\pi^{-}}\ . 
        \end{equation}  
The above considerations are easily extendable to fermionic fields and to
$AdS_4$ and $dS_4$ foliations. One additional rule, which should be followed, is that in the passage from  
orbifold to two intervals the $\epsilon(y)$ becomes decomposed as  
$\epsilon(y) = H_{-}(y) + H_{+}(y)$ where $H_{-}(y) = -1 \;\; {\rm if} \;\; 
y < 0 \;\; {\rm and} \;\; 0 \;\;{\rm otherwise}$, and $H_{+}(y) =  1 \;\; {\rm if} \;\; 
y > 0 \;\; {\rm and} \;\; 0 \;\;{\rm otherwise}$. 
\section{Consistency conditions in the interval picture} 
For a 5d metric of the form   
        \begin{equation} \label{metrykadlareg} 
        ds^{2}=e^{2A(y)}\bar{g}_{\mu\nu}dx^{\mu}dx^{\nu}+dy^{2}\  
        \end{equation}  
one can derive, using equations of motion, consistency conditions \cite{Forste:2000ps,Forste:2000ft} 
-- 
relations between energy-momentum tensor and metric tensor, sometimes called
sum rules \cite{sumrules}:  
        \begin{equation} \label{regsum}  
        \left(A^{\prime}e^{nA}\right)^{\prime}=\frac{1}{12M^{3}}e^{nA}\left(T_{\mu}^{\mu}+(2n-4)T_{5}^{5}\right)-\frac{1-n}{12}e^{n-2}\bar{R}\ , 
        \end{equation}  
        where $n $ is an arbitrary number and $\bar{R}$ denotes the 4-dimensional curvature scalar. 
As an application of the orbifold picture--interval picture correspondence that we have established, we shall formulate these sum rules in the interval picture.  
 
The energy-momentum tensor derived from the orbifold action  
 is 
	\begin{equation}  
	T^{TOT}_{\mu\nu}=T_{\mu\nu}+T^{0}_{\mu\nu}\delta(y)+T^{\pi}_{\mu\nu}\delta(y-\pi r_{c})\ ,\qquad T^{TOT}_{55}=T_{55}\ . 
	\end{equation}  
After integrating (\ref{regsum}) over the orbifold we obtain
	\begin{eqnarray} \label{regsumkon}&& \oint_{-\pi r_{c}}^{\pi r_{c}}\left(\frac{1}{12M^{3}}e^{nA}\left(T_{\mu}^{\mu}+(2n-4)T_{5}^{5}\right)-\frac{1-n}{12}e^{n-2}\bar{R}\right)dy=  \nonumber \\ & =&-\frac{1}{12M^{3}}\left(e^{nA(0)}(T^{0})_{\mu}^{\mu}+e^{nA(\pi r_{c})}(T^{\pi})_{\mu}^{\mu}\right)\ .  \end{eqnarray}  
\vskip0.3cm 
Let us consider the above set-up on a pair of intervals.  
In this case, the energy-momentum tensor does not include the $T^{0^{\pm}}_{\mu\nu}$ and $T^{\pi^{\pm}}_{\mu\nu}$ contributions. Instead, we have additional boundary equations (\ref{warbrz1}).  
Equation (\ref{regsum}), integrated over the internal space, leads to  
	\begin{eqnarray}\label{regsumdlaint} &&  \int_{-\pi r_{c}}^{\pi r_{c}}\left(\frac{1}{12M^{3}}e^{nA}\left(T_{\mu}^{\mu}+(2n-4)T_{5}^{5}\right)-\frac{1-n}{12}e^{n-2}\bar{R}\right)dy= \nonumber \\ &  =&A^{\prime}e^{nA}\Big|_{y=\pi r_{c}^{-}}-A^{\prime}e^{nA}\Big|_{y=-\pi r_{c}^{+}}-A^{\prime}e^{nA}\Big|_{y=0^{+}}+A^{\prime}e^{nA}\Big|_{y=0^{-}}\ . \end{eqnarray} 
The tensor $W^{5}_{MN}$ calculated for the metric (\ref{metrykadlareg}) reads 
	\begin{eqnarray} &&   W^{5}_{\mu\nu}=-3A^{\prime}e^{2A}\bar{g}_{\mu\nu}\ , \nonumber \\ && W^{5}_{55}=0\ .  \end{eqnarray} 
After taking equation (\ref{warbrz1}) into account, we obtain 
       \begin{eqnarray} &&  A^{\prime}\Big|_{y=0^{+}}=\frac{(T^{0^{+}})^{\mu}_{\mu}}{24M^{3}}\ ,\qquad  A^{\prime}\Big|_{y=\pi r_{c}^{-}}=-\frac{(T^{\pi^{+}})^{\mu}_{\mu}}{24M^{3}}\ ,  \\  &&  A^{\prime}\Big|_{y=0^{-}}=-\frac{(T^{0^{-}})^{\mu}_{\mu}}{24M^{3}}\ ,\qquad A^{\prime}\Big|_{y=-\pi r_{c}^{+}}=\frac{(T^{\pi^{-}})^{\mu}_{\mu}}{24M^{3}}\ ,\end{eqnarray}  
        and eventually (\ref{regsumdlaint}) takes the form 
        \begin{eqnarray} &&  \int_{-\pi r_{c}}^{\pi r_{c}}\left(\frac{1}{12M^{3}}e^{nA}\left(T_{\mu}^{\mu}+(2n-4)T_{5}^{5}\right)-\frac{1-n}{12}e^{n-2}\bar{R}\right)dy= \nonumber \\ &  =&-\frac{1}{24M^{3}}\left((T^{\pi^{+}})^{\mu}_{\mu}e^{nA(\pi r_{c}^{-})}+(T^{\pi^{-}})^{\mu}_{\mu}e^{nA(-\pi r_{c}^{+})}+(T^{0^{+}})^{\mu}_{\mu}e^{nA(0^{+})}+(T^{0^{-}})^{\mu}_{\mu}e^{nA(0^{-})}\right)\ .  \end{eqnarray} 
This is fully equivalent to equation (\ref{regsumkon}) when the  
conditions (\ref{warunkirownowa}) are satisfied. 
\vskip0.5cm 

It has been established in \cite{Forste:2000ps,Forste:2000ft}
that consistency of five-dimensional solutions with naked singularities \cite{Arkani-Hamed:2000eq,kss}
requires the insertion of finely tuned sources at the positions of singularities. 
The usual way to arrive at such solutions is to start with an infinite space with vanishing boundary conditions. We solve equations of motion for all fields and arrive at a solution with naked singularity at a position $y_s$. We then notice that after inserting additional sources we can obtain a solution that is valid in the whole infinite space by taking all fields to vanish beyond $y_s$.   
Alternatively, we can continue fields periodically for $y >y_s$. Let us demonstrate how the same physical situation is reached when one starts 
with the consistent action on the interval:  
	\begin{eqnarray}\label{kssdzialanie} & S&=M^{3}\int d^{4}x\int_{0}^{L} dy\sqrt{-G}\left(R-\frac{4}{3}G^{MN}\partial_{M}\Phi\partial_{N}\Phi\right) -M^{3}\int d^{4}x\sqrt{-g^{0}}V_{0}e^{b_{0}\Phi}\Big|_{y=0}+ \nonumber \\ & & -M^{3}\int d^{4}x\sqrt{-g^{L}}V_{L} e^{b_{L}\Phi}\Big|_{y=L}+M^{3}\int d^{4}x\sqrt{-G}\bar{W}^{5}\Big|_{y=0}-M^{3}\int d^{4}x\sqrt{-G}\bar{W}^{5}\Big|_{y=L}  \end{eqnarray} 
(we have included Gibbons--Hawking terms). 
The variation with respect to the metric tensor (here and below we follow the  
procedure given earlier, i.e. we allow for non-vanishing boundary variations)  
leads to the Einstein equation in the bulk 
        and to the conditions 
	\begin{equation}  \label{warunkigrawkss} 
	\sqrt{-G}W^{5}_{\mu\nu}\Big|_{y=0}=\sqrt{-g^{0}}\frac{1}{2}V_{0}e^{b_{0}\Phi}g^{0}_{\mu\nu}\Big|_{y=0},\quad \sqrt{-G}W^{5}_{\mu\nu}\Big|_{y=L}=-\sqrt{-g^{L}}\frac{1}{2}V_{L}e^{b_{L}\Phi}g^{L}_{\mu\nu}\Big|_{y=L} 
	\end{equation}  
        on the boundary. 
The variation with respect to the scalar field gives  
the standard equations of motion in the bulk together with the boundary conditions 
	\begin{equation} \label{warunkiskalarnewkss} 
	\sqrt{-G}\frac{8}{3}G^{55}\partial_{5}\Phi\Big|_{y=0}=-\sqrt{-g^{0}}b_{0}V_{0}e^{b_{0}\Phi}\Big|_{y=0}\ ,\quad\sqrt{-G}\frac{8}{3}G^{55}\partial_{5}\Phi\Big|_{y=L}=\sqrt{-g^{L}}b_{L}V_{L}e^{b_{L}\Phi}\Big|_{y=L}\ . 
	\end{equation}  ) 
This problem has the following singular solution known from \cite{kss} (the warp factor is $e^{A(y)}$): 
	\begin{equation}  
	\Phi(y)=-\frac{3}{4} \log\left(c-\frac{4}{3}y\right)+d\ ,\quad A(y)=\frac{3}{4} \log\left(c-\frac{4}{3}y\right)+d_{A}\ , 
	\end{equation}  
       where  $c$, $d$, $d_{A}$ are integration constants.  
Since this solution has a singularity at  
 $y=\frac{3}{4}c$, we should take $L=\frac{3}{4}c$. One can see that this causes no problem in the boundary conditions  
 (\ref{warunkigrawkss}) and (\ref{warunkiskalarnewkss}). This is so  because the boundary terms in (\ref{kssdzialanie}) remain finite at  
the position of the singularity.  
One may suspect that the regularity of this solution is somehow related to  
the fact that the vacuum action, without contributions the boundaries, is 
non-vanishing, but finite. To clarify this issue let us take a model, where  
the vacuum action is infinite (by vacuum action we mean the vacuum Lagrangian integrated over $y$ only). This is for instance the case for effective action of the nonsupersymmetric  
Type I string  given in \cite{dudas}. This action in the Einstein frame reads 
        \begin{equation} \label{dudasdzialanie} 
        S_{E}=\frac{1}{2k^{2}}\int d^{10}x\sqrt{-G}\left(R-\frac{1}{2}(\partial\Phi)^{2}-2\alpha_{E}e^{\frac{3}{2}\Phi}\right)\ .        
        \end{equation}  
        Classical background preserving a nine-dimensional Poincare symmetry  
has the form 
        \begin{equation}  
        ds^{2}=e^{2A(y)}\eta_{\mu\nu}dx^{\mu}dx^{\nu}+e^{2B(y)}dy^{2}\ ,\qquad\Phi=\Phi(y)\ , 
        \end{equation}  
        where $\mu,\nu=0,...,8$ and $y$ denotes the transverse coordinate. 
The solution in the Einstein frame is 
        \begin{eqnarray}\label{dudasrozw} & \Phi=\frac{3}{4}\alpha_{E}y^{2}+\frac{2}{3}\log|\sqrt{\alpha_{E}}y|+\Phi_{0}\ , & \nonumber \\ &   ds^{2}_{E}=|\sqrt{\alpha_{E}}y|^{1/9}e^{-\alpha_{E}y^{2}/8}\eta_{\mu\nu}dx^{\mu}dx^{\nu}+|\sqrt{\alpha_{E}}y|^{-1}e^{-9\alpha_{E}y^{2}/8}e^{-3\Phi_{0}/2}dy^{2}\ .& \end{eqnarray} 
        Notice that this solution has  singularities at $y=0$ and $y=\infty$.  
Hence, it is natural to restrict the transverse coordinate to the region  
$0<y<\infty$. 
The naively calculated cosmological constant is 
        \begin{equation}   
        \Lambda_{eff}=\frac{1}{2k^{2}}\int_{0}^{\infty} e^{\frac{3}{4}\Phi_{0}}\alpha_{E}^{3/2}\frac{y}{2}dy=\frac{1}{2k^{2}} e^{\frac{3}{4}\Phi_{0}}\alpha_{E}^{3/2}\frac{y^{2}}{4}\Bigg|_{0}^{\infty}\ , 
        \end{equation}  
which is infinite.  
This points towards some inconsistency: 9d Poincare invariance  
is maintained, but naive 9d cosmological constant is infinite.  
Let us try to repair the system following the prescription formulated 
earlier. We add to the action (\ref{dudasdzialanie}) the Hawking--Gibbons terms 
        \begin{equation} \label{dudashg} 
        S_{HG}=-\frac{1}{2k^{2}}\int d^{9}x \sqrt{-G}\bar{W}^{10}\Bigg|_{0}^{\infty}\ ,  
        \end{equation}  
and, in addition, we augment it by boundary branes  
        \begin{equation} \label{dudasbrany} 
        S_{br}=\frac{1}{2k^{2}}\int d^{9}x \sqrt{-g^{0}}V_{0}(\Phi)\Bigg|_{y=0}+\frac{1}{2k^{2}}\int d^{9}x \sqrt{-g^{\infty}}V_{\infty}(\Phi)\Bigg|_{y=\infty}\ , 
        \end{equation}  
with tensions determined by the conditions         
	\begin{equation} \label{plim}  
	\sqrt{-G}W^{10}_{\mu\nu}\Big|_{y=0}=-\sqrt{-g^{0}}\frac{1}{2}V_{0}(\Phi)g^{0}_{\mu\nu}\Bigr|_{y=0}\ ,\quad\sqrt{-G}W^{10}_{\mu\nu}\Big|_{y=\infty}=\sqrt{-g^{\infty}}\frac{1}{2}V_{\infty}(\Phi)g^{\infty}_{\mu\nu}\Bigr|_{y=\infty}\ . 
	\end{equation}  
        Adding the brane at infinity may seem questionable; however, it should be noticed that the volume of the transverse dimension, 
        \begin{equation}  
        L_{10}=\int_{0}^{\infty}dy e^{B(y)}=e^{-3\Phi_{0}/4}\alpha_{E}^{-\frac{1}{4}}\int_{0}^{\infty}\frac{dy}{y^{1/3}}e^{-9\alpha_{E}y^{2}/16}\ , 
        \end{equation}  
is finite. Hence, when one goes over to the coordinate that is proportional to the proper volume of that dimension, the position of the second singularity becomes finite as well.  
 
Let us check how the boundary terms modify the 9d cosmological constant. 
After inserting the vacuum solutions into         
(\ref{dudashg}) and (\ref{dudasbrany}) we obtain 
        \begin{equation}  
        S_{HG}^{eff}=-\frac{1}{2k^{2}} e^{\frac{3}{4}\Phi_{0}}\alpha_{E}^{3/2}\frac{9y^{2}}{4}\Bigg|_{0}^{\infty}\ ,\quad S_{br}^{eff}=\frac{1}{2k^{2}} e^{\frac{3}{4}\Phi_{0}}\alpha_{E}^{3/2}\frac{8y^{2}}{4}\Bigg|_{0}^{\infty}\ . 
        \end{equation}  
Thus the final result vanishes, 
        \begin{equation}  
        \Lambda_{eff}^{\prime}=\Lambda_{eff}+S_{HG}^{eff}+S_{br}^{eff}=0\ , 
        \end{equation}  
as it should be.  
This example has a number of interesting features. Firstly, we note that  
brane tensions at the endpoints can easily be computed with the help of (\ref{plim}); the tension $V_0$   vanishes, whereas the second one,  
$V_{\infty}$, becomes  
infinite. One might wonder whether a brane with an infinite tension may arise  
within a physical system. One way to make sense of such a situation is to  
view the tension $V_{\infty}$ not as an independent parameter of the  
system, but rather as a quantity adjusting itself, through equation (\ref{plim}), to the dynamics of the rest of the system. This is another face of the  
fine-tuning identified in \cite{Forste:2000ps}.  
Secondly, one can test the infinite tension brane by computing the acceleration of a freely falling test particle in its vicinity. It turns out that such a particle accelerates towards the infinite tension brane (and the acceleration becomes infinite on the brane). The second brane, at $y=0$, attracts test particles as well, but this time the acceleration vanishes on the brane.  
This behaviour corresponds to the case of a positive mass  
Schwarzchild black hole, thus indicating that the branes at singularities  
are physical in the present case \cite{Maldacena:2001mw}.  
Further, we note that the acceleration vanishes at a special point between branes: $y_{a}=\frac{2}{3 \sqrt{\alpha}}$ (we note that $A'(y_{a})=0$). It
is therefore natural to propose to 
regularize the system by defining a thick brane, which corresponds to integrating the 10d Lagrangian between $y_{a}$ and $y=\infty$. This can easily be done, 
using the formulae derived earlier and the explicit form of the vacuum solution. The effective brane tension $V_{a, \, eff}$ is found to be zero.  
This is the same as for a thin brane put at the position $y_a$, with its brane tension determined directly  from the jump conditions (\ref{plim}).  
Thus the infinite tension brane can be made a `physical' object through  
smearing it over a finite distance along the transverse dimension, and this  
turns out to be equivalent to putting a thin regulator brane in front of the  
singularity.  
\vskip0.3cm 
\section{Summary} 
In this note we describe the role of boundary terms when passing from the  
orbifold description of a brane world to the interval description.  
We have noticed that it is natural, in the interval picture, to use  
field variations that are non-vanishing at the endpoints. This, although  
equivalent to results obtained with any other form of the variational principle, gives rise to the observation that boundary terms can be seen as related  
to certain bulk operators through equations of motion.  
We discuss the case of $Z_2$-even and $Z_2$-odd orbifold scalars and  
that of the metric tensor.  
The relations between orbifold and interval pictures, described here for  
scalar fields, can be extended to fermionic fields as well. \\ 
As one of the applications of the interval picture, we reconsider  
models with naked singularities. In particular, we discuss the situation, where the naive $(d-1)$-dimensional cosmological constant appears to be infinite. 
We show that proper inclusion of boundary terms makes the solution consistent, 
and in particular makes the effective cosmological constant vanish.  
Moreover, we argue that sense can be made of an infinite tension brane,  
by making it a finite width brane, which turns out to be equivalent to  
putting into a system a thin shielding brane in front of the troublesome  
singularity.   
\vskip0.7cm
The work of the authors has been supported by RTN programs HPRN-CT-2000-00152 and HPRN-CT-2000-00148, and by the Polish Committee for Scientific Research 
under grants
5~P03B~119 20 (2001-2002) (R.M.) and 5~P03B~150 20 (2001-2002) (Z.L.).
 
\end{document}